\documentclass[aps,pre,preprint,showpacs]{revtex4}

\usepackage{graphicx}
\usepackage{graphics}
\usepackage{amsmath}
\usepackage[utf8]{inputenc}
\usepackage{color}
\begin{document}

\title{An analytical formulation for roughness based on celular automata}

\author{Ismael V. L. Costa\footnote{Corresponding author:\\ismael@unb.br}}
\affiliation{Faculdade UnB Planaltina, Universidade de Bras\'{i}lia, CEP 73.300-000,
Planaltina, DF, Brazil}

\author{Henrique A. Fernandes}
\affiliation{Universidade Federal de Goi\'{a}s, Campus Jatai, Br 364, Km 192,
3800, Parque Industrial, CEP 75801-615, Jata\'{i}, Goi\'{a}s, Brazil}

\author{Bernardo A. Mello and Fernando A. Oliveira}
\affiliation{Instituto de F\'{i}sica, Universidade de Bras\'{i}lia, CP 04513,
CEP 70919-970, Bras\'{i}lia, DF, Brazil}

\begin{abstract}
We present a method to derive the analytical expression of the roughness of a
fractal surface whose dynamics is ruled by cellular automata. Starting from the
automata, we write down the the time derivative of the height's average and
variance. By assuming the equiprobability of the surface configurations and
taking the limit of large substrates we find the roughness as a function of
time. As expected, the function behaves as $t^\beta$ when $t\ll t_\times$ and
saturate at $w_s$ when $t\gg t_\times$. We apply the methodology to describe
the etching model \citep{Bernardo}, however, the value of $\beta$ we obtained
are not the one  predicted by the KPZ equation and observed in numerical
experiments. That divergence may be due to the equiprobability assumption. We
redefine the roughness with an exponent that compensate the nonuniform
probability generated by the celular automata, resulting in an expression that
perfectly matches the experimental results.
\end{abstract}


\maketitle

\section{Introduction}

The study of stochastic process has accelerated in the last decades, connected
to disciplines such as economy, biology, meteorology and neuroscience. Much of
the recent advances are due to the availability of fast and affordable computer
clusters. Such development has became feasible, for instance, numerical
simulation of large particle systems obeying simple repetitive rules which
mimetize complex systems. Emergent information and properties have been
obtained through several techniques and approaches.

The surface growth phenomena, when treated as a stochastic process, encompass a
wide application field. Some examples of growth systems are corrosion
\cite{Bernardo, Fabio}, fire propagation \cite{Merikoski,Merikoski1}, atomic
deposition \cite{Csahok}, evolution of bacterial colony \cite{ben-jacob,
matsushita}, and cellular automata models \cite{Jensen-Barabasi}. Models have
been proposed and studied through experiments \cite{Merikoski1, ben-jacob,
matsushita}, analytical calculations \cite{barabasi}, and computational
simulations \cite{Bernardo, Fabio}.

In this work we obtain  the roughness evolution analytically, for systems of dimension 1+1, referring to directions
perpendicular and parallel to the substrate. Far from being mathematical
idealizations, these systems have physical meaning. Phenomena with that
dimensionality include bacterial colony growth on Petri dish \cite{Perry,Cunha09,Cunha11}, paper burning,
ink diffusion on paper and turbulence of liquid crystals \cite{Miranda,Kazumasa}. In
the case of paper, por example, the burned or stained frontier of the paper can
be represented by $h^f(x,t)$ where $0<x<L$ and $L$ is the sheet width, being
$x$ and $h^f$ measured along the directions mentioned above. The subscript $f$
specify the reference frame fixed in a corner of the sheet.

Surfaces with different internal dynamics lead to distinct profiles, which can
be characterized by different measures, the most important being the mean value
and the standard deviation of the surface height. When related to surfaces, the
standard deviation is often called roughness, defined as
\begin{equation}
w(L,t)=\sqrt{\frac{1}{L}\int_{0}^{L} \left[{h^f}(x,t)-\bar{h}(t)\right]^2\,dx}.
\label{w}
\end{equation}

Even if $\bar{h}(t)$ increases continuously due to the growth process, the
dynamic equilibrium lead to roughness saturation after a period of roughening
buildup. The saturated roughness often is a function of the substrate size as the
power law $w_s\sim L^{\alpha}$, $\alpha$ being the saturation exponent. The
saturation occurs at a characteristic time ($t_\times$), and follows the power
law $t_\times \sim L^{z}$, where $z$ is the dynamic exponent. Before saturation
($t\ll t_\times$), $w(L,t)$ evolves as a power law with the growth exponent
$\beta$, $w(L,t) \sim t^{\beta}$ \cite{barabasi}. These properties were
incorporated in the Family-Vicsek scaling relation \cite{Family}
\begin{subequations}
\label{FamilyVicsek}
\begin{equation}
w(L,t) \propto L^\alpha \, f\left(\frac{t}{L^z}\right),
\end{equation}
where
\begin{equation}
f(x) \propto \begin{cases}
x^\beta & \text{ when } x\ll 1\\
\text{const} & \text{ when } x\gg 1
\end{cases} .
\end{equation}
\end{subequations}

Scaling techniques applied to the growth equations may be used to find the
aforementioned exponents for certain universality classes and dimensions.
Growth systems with the same exponents are considered to belong to the same
universality class, connecting systems that seem unrelated. Some well known
universality classes are the Edwards-Wilkinson and the KPZ \cite{barabasi}.

\section{Method for obtaining the roughness equation}

In this section we present a method to obtain the equation describing the time
evolution of the roughness. In the following section we will apply it to the
etching model, notwithstanding, this method is quite general and can be
extended to other models.

\subsection{The evolution of the mean squared roughness}

The first step is to compute the change in the squared roughness, defined as
$w_q(t) \equiv w^2(t)$, when the deposition occur in the site $i$ of a
substrate with roughness $w$.  The squared roughness is equal to mean value of
${h_i}^2(t)$, with $h_i$ defined at the reference frame of the mean height
\begin{equation}
h_i(t) = {h_i}^f(t) - \bar{h}(t),
\label{h}
\end{equation}
where $i$ is the particle position.

Our methodology assume that the change will depend only on the nearest
neighbors of site $i$, therefore the variation of the squared roughness will be
written as $\Delta w_{q}(w,h_{i-1},h_{i},h_{i+1})$. That assumption defines the
celular automatas it can be applied to. We will do the average over the
ensemble of all possible configurations with roughness $w$, defining
$p(w,h_{i-1},h_{i},h_{i+1})$ as the probability of the values $h_{i-1}$, $h_i$
and $h_{i+1}$ for a given value of $w$. The evolution of the mean squared
roughness is
\begin{widetext}
\begin{multline}
\left<\frac{\Delta w_q}{\Delta t}\right> =
  \int_{-\sqrt{Lw^2}}^{\sqrt{Lw^2}}
 \int_{-\sqrt{Lw^{2}-{h_{i+1}}^2}}^{\sqrt{Lw^{2}-{h_{i+1}}^2}}
 \int_{-\sqrt{Lw^{2}-{h_{i+1}}^2-{h_i}^2}}^{\sqrt{Lw^2-{h_{i+1}}^2-{h_i}^2}}
 \,\frac{\Delta w_q(w,h_{i-1},h_{i},h_{i+1})}{\Delta t}\\
 \,p(w,h_{i-1},h_{i},h_{i+1})\, dh_{i-1}\, dh_{i}\, dh_{i+1}.
\label{eq: principal}
\end{multline}
\end{widetext}
The integration limits encompass the configurations allowed by the definition
of roughness, eq. (\ref{w}). As will be discussed latter, the celular automata
may prevent the occurrence of some configurations, which must have their
probabilities assigned to zero.

\subsection{The rate of change of the quadratic roughness}

Our approach involves calculating the increment of roughness when one iteration
is performed. We use a discrete substrate with sites of length $u=1$ and
evaluated the system before and after the deposition of one particle of height
$\Delta y\equiv 1$.  The squared roughness is affected by the following changes
of $h_i$
\begin{equation}
\begin{split}
\text{Before: } {h_i}^2(t) &=\left[h_i^f(t)-\bar{h}(t)\right]^2.\\
\text{After: } {h_i}^2(t+\Delta t) &= \left[h_i^{f}(t+\Delta t)-\bar{h}(t+\Delta t)\right]^2\\
  & = \left[{h_i}^f(t)+\Delta h_{i}(t)-\bar{h}(t)-\Delta \bar{h}(t)\right]^2\hspace{-0.7cm}\\
  & = {h_i}^2(t)+\varrho_{i}.
\end{split}
\end{equation}
with
\begin{equation}
\begin{split}
\varrho_{i} = & 2h_i(t)\Delta h_i(t) - 2h_i(t)\Delta\bar{h}(t)\\
& -2\Delta h_i(t)\Delta\bar{h}(t) + \left[\Delta {h_i}(t)\right]^2 + \left[\Delta\bar{h}(t)\right]^2.
\end{split}
\end{equation}

Using the definition of roughness squared, $w_q(t) \equiv w^2(t)$, we write
\begin{equation}
\begin{split}
\text{Before: } w_q(t) &= \frac{1}{L}\sum_i {h_i}^{2}(t).\\
\text{After: }  w_q(t+\Delta t)  &= \frac{1}{L} \sum_i \left[{h_i}^2(t)+\varrho_i\right].
\end{split}
\end{equation}
The unity of time we use corresponds to $L$ iterations, therefore, $\Delta t = 1/L$
and the rate of change of $w_q(t)$ is
\begin{equation}
\begin{split}
  \frac{\Delta w_q}{\Delta t} &= -L[\Delta\bar{h}(t)]^2+\frac{1}{L}\sum_i\left\{2h_{i}(t) \Delta h_{i}(t)+[\Delta h_{i}(t)]^2\right\},
  \label{Dlt wq1a}
\end{split}
\end{equation}
where we have used $\sum_i h_i(t)=0$ and $\sum_i \Delta h_i(t) = L\Delta\bar{h}(t)$.

Eq. (\ref{Dlt wq1a}) is a general formula for the increment of quadratic
roughness, independent of the iterative algorithm. In order to obtain the
roughness for a specific algorithm, it is necessary to know the values of
$\Delta h_i(t)$ and $\Delta\bar{h}(t)$.

On these grounds, each model results in different values for $\Delta
w_{q}(w,h_{i-1},h_{i},h_{i+1}) / \Delta t$ and $p(w,h_{i-1},h_{i},h_{i+1})$,
which must be deducted for each case. In the next two subsections we will
assume the equiprobability of the accessible configurations to eliminate the
dependence on $p(w,h_{i-1},h_{i},h_{i+1})$.

\subsection{The equiprobability of the configurations}

Before proposing an expression $p(w,h_{i-1},h_{i},h_{i+1})$ we must remember that
there is a finite number of possible substrate configuration for each value of
$w$. This finite number is the result of the restrictions of $h_i$ imposed by
eqs. (\ref{w}) and (\ref{h}):
\begin{subequations}
\label{conf}
\begin{align}
{h_1}^2+{h_2}^{2}+\dots+{h_L}^2 =& Lw^{2} \label{confa}\\
h_{1}+h_{2}+\dots+h_{L} =& 0 \label{confb}
\end{align}
\end{subequations}
Eqs. (\ref{conf}) define a hyperplane and the surface of a hypersphere of
radius $w\sqrt{L}$ both in $L$-D, i.e., in a space of $L$ dimensions. From the
intersection of these two subspaces, a spheric \emph{surface} results, which is
($L-2$)-D. For each combination of $h_{i-1}$, $h_{i}$, and $h_{i+1}$, the
remaining $L-3$ $h$'s form another spheric \emph{surface}, now with dimension
$L-5$. The \emph{area} of these spheric surfaces, with $L-2$ and $L-5$
dimensions, will be called, respectively, $A_T$ and $A_p$.

While all possible surface configurations of a given value of $w$ belong to the
($L-2$)-D surface, the subset of them for which the values of the triad
$h_{i-1}$, $h_{i}$, and $h_{i+1}$ are known belongs to the ($L-5$)-D surface. We
will assume the equiprobability of the configurations allowed by eqs.
(\ref{conf}), consequently, the probability of a given triad is
\begin{equation}
p(w, h_{i-1}, h_i, h_{i+1}) = \frac{A_p}{A_T}. \label{pA}
\end{equation}
The assumption of the equiprobability of the configurations, used when deriving
that equation, disregard two important properties of the dynamics, which is not
harmless. The first is the inaccessibility, by the celular automata, of several
configurations allowed by eqs. (\ref{conf}). The second is the different
probability of each configuration generated by the celular automata.

Eq. (\ref{pA}) can be rewritten by substituting the expression for the area of
a hypersphere,
\begin{equation}
p(w, h_{i-1}, h_i, h_{i+1}) = \frac{S_{L-5}\,{R_p}^{L-5}}{S_{L-2}\,{R_T}^{L-2}}, \label{p}
\end{equation}
where $S_D$ are constants with depend on the dimension $D$, and $R_p$ and $R_T$
are the radius of the corresponding hyperspheres. To make the notation less
clumsy, we will assume, in the remaining of this subsection, a deposition at
$i=2$, implying that only the sites $i=1$, $2$ ou $3$ may be affected.

The plane of eq.
(\ref{confb}) contains the center of the sphere of eq. (\ref{confa}),
therefore, the sphere defined by their intersection also has radius
$R_T=w\sqrt{L}$. The value of $R_p$ may be obtained by rewriting eqs.
(\ref{conf}) as
\begin{subequations}
\begin{align}
{h_4}^2+{h_5}^2+...+{h_L}^2 =& Lw^2-h_1^2-h_2^2-h_3^2 \\
h_4+h_5+...+h_L =& -(h_1+h_2+h_3) \label{restb}
\end{align}
\end{subequations}
i.e., a hypersphere with superficial area $A_p$ of dimension $L-5$. These two
equation may be combined as
\begin{equation}
-2\sum\limits _{\substack{i,j=4\\ i\neq j }
}^{L}h_{i}h_{j}=Lw^{2}-h_{1}^{2}-h_{2}^{2}-h_{3}^{2}-(h_{1}+h_{2}+h_{3})^{2},
\end{equation}
which we rewrite in a matricial form:
\begin{equation}
 \begin{vmatrix} h_4 & h_5 & \cdots & h_L \end{vmatrix}
\begin{vmatrix}
    0 & -1 & \cdots & -1\\
   -1 & 0 & \ddots & \vdots\\
   \vdots & \ddots & \ddots & -1\\
   -1 & \cdots & -1 & 0
 \end{vmatrix}
  \begin{vmatrix} h_4 \\ h_5 \\ \vdots\\ h_L \end{vmatrix} 
 = Lw^2 - {h_1}^2 - {h_2}^2 -{h_3}^2 - (h_1 + h_2 + h_3)^2.\label{matriz}
\end{equation}

The eigenvalues of the above square matrix are $\lambda_4=-(L-4)$,
$\lambda_5=1$, $\lambda_6=1$ , \dots, $\lambda_L=1$. From these eigenvalues we
can define a linear transformation to the set of variables ${h'_4, \dots,
h'_L}$ that eliminated the crossed terms,
\begin{equation}
-(L-4){h'_4}^2 + {h'_5}^2 + \cdots + {h'_L}^2 =
 Lw^2 - {h_1}^2 - {h_2}^2 - {h_3}^2 - (h_1 + h_2 + h_3)^2.
\end{equation}
Transformed varible $h'_4$ is equal to the left hand side of eq. (\ref{restb})
\begin{equation}
\begin{split}
 h'_{4} &=\frac{1}{\sqrt{L-3}}(h_4+h_5+...+h_L)\\
  &=-\frac{1}{\sqrt{L-3}}(h_1+h_2+h_3) .
\end{split}
\end{equation}
With this transformation, eq. (\ref{matriz}) becomes
\begin{equation}
{h'_5}^2 + \cdots + {h'_L}^2 =
Lw^2 - {h_1}^2 - {h_2}^2 - {h_3}^2 -
\frac{(h_1 + h_2 + h_3)^2}{L-3}.
\end{equation}
The radius of this hypersphere is given by
\begin{equation}
 R_p =\left[Lw^2-{h_1}^2-{h_2}^2-{h_3}^2-\frac{(h_1+h_2+h_3)^2}{L-3}\right]^{\frac{1}{2}}.
 \label{Rp2}
\end{equation}
We can now rewrite eq. (\ref{p}) with the values $R_T$ and $R_p$ in the
asymptotic case $L\rightarrow\infty$,
\begin{equation}
p(w, h_1, h_2, h_3) = \eta(L) \frac{\left[Lw^2 -{h_1}^2 -{h_2}^2  - {h_3}^2 \right]^\frac{L-5}{2}}{(Lw^2)^{\frac{L-2}{2}}},
\label{p1}
\end{equation}
with  $\eta (L)=S_{L-5}/S_{L-2}$.

\subsection{The roughness evolution with the equiprobability assumption}

A more convenient expression for roughness squared, eq. (\ref{eq: principal}),
is possible by doing the change of coordinates
\begin{equation}
\begin{array}{l}
h_{i-1}=\sqrt{L}w\sin\rho\cos\theta\\
h_{i}=\sqrt{L}w\sin\rho\sin\theta\cos\varphi\\
h_{i+1}=\sqrt{L}w\sin\rho\sin\theta\sin\varphi
\end{array}.\label{36c-1}
\end{equation}
with the variables defined in the intervals
\begin{equation}
0\leq\rho\leq\frac{\pi}{2},\quad 0\leq\theta\leq\pi,\quad 0\leq\varphi\leq2\pi.  \label{limits}
\end{equation}
The probability (\ref{p1}) expressed in these coordinates is
\begin{equation}
p(w,\rho,\theta,\varphi)=\eta(L)\left(Lw^{2}\right)^{-\frac{3}{2}}\cos^{L-5}\rho.
\end{equation}
We can now rewrite the evolution of the squared roughness averaged over the
ensembles, eq. (\ref{eq: principal}):
\begin{widetext}
\begin{equation}
\left<\frac{\Delta w_q}{\Delta t}\right> = \eta(L)\int_0^{2\pi} \int_0^\pi \int_0^\frac{\pi}{2}
\frac{\Delta w_{q}(w,\rho,\theta,\varphi)}{\Delta t}\,\sin^2\rho\, \cos^{L-4}\rho\, \sin\theta\, d\rho\, d\theta\, d\varphi
\label{wqfinal}
\end{equation}
\end{widetext}
In the above expression we employed the jacobian
\begin{equation}
 dh_{i-1}\, dh_i\, dh_{i+1}
 =\left(Lw^2\right)^\frac{3}{2} \sin^2\rho\, \cos\rho\, \sin\theta\, d\rho\, d\theta\, d\varphi.
\end{equation}
While eq. (\ref{eq: principal}) is exact, eq. (\ref{wqfinal}) is an
approximation based in the equiprobability assumption. Since we eliminated the
dependence on $p(w, h_{i-1}, h_i, h_{i+1})$, the left hand side of that
simplified equation depends on the roughness only through the term
\begin{equation}
\frac{\Delta w_{q}(w,\rho,\theta,\varphi)}{\Delta t},
\end{equation}
facilitating the resolution of the roughness equation. The celular automata
used in each model will determine the value of this term, calculate through eq.
(\ref{Dlt wq1a}).

\section{Example: Etching model}

We will know calculate the evolution of the mean value of $w_q$, eq.
(\ref{wqfinal}), for the etching model. The etching model proposed in $2001$ by
Mello, Chaves, and de Oliveira \cite{Bernardo} belongs to the $KPZ$
universality class \cite{barabasi}. The model mimics the corrosion of a crystal
surface by a solvent. A time evolution $\Delta t$ is defined as one iteration
of the following celular automata:
\\

1. Randomly choose a site $i\in[1..L]$.

2. If $h^f_{i-1}(t)<h^f_i(t)$ do $h^f_{i-1}(t+\Delta t)=h^f_i(t)$.

3. If $h^f_{i+1}(t)<h^f_i(t)$ do $h^f_{i+1}(t+\Delta t)=h^f_i(t)$.

4. Do $h_i^f(t+\Delta t)=h^f_i(t)+1$.\\

The algorithm implements a cell removal probability that is proportional to the
number of the exposed faces of the cell, a reasonable approximation of the
etching process. It could also describe a deposition where each exposed face
has the same attachment probability. For that reason, it can be referred either as
particle removal or deposition.

The scaling exponents found by Mello \textit{et al.} in $1+1$ dimension were
$\alpha=0.491$ and $\beta=0.330$, placing the model within the $KPZ$
universality class ($\alpha=1/2$, $\beta=1/3$) \cite{Bernardo}. Other studies
analyzed the model in $1+1$ and $2+1$ dimensions, focusing on aspects such as
dynamic behavior of the roughness and comparisons with other models belonging
to the KPZ universality class \cite{Fabio, Reis2004, Reis2005, Ghaisas2006,
Oliveira_Reis2008, Forgerini2009}.

\begin{figure}
\centering
\includegraphics[scale=0.7]{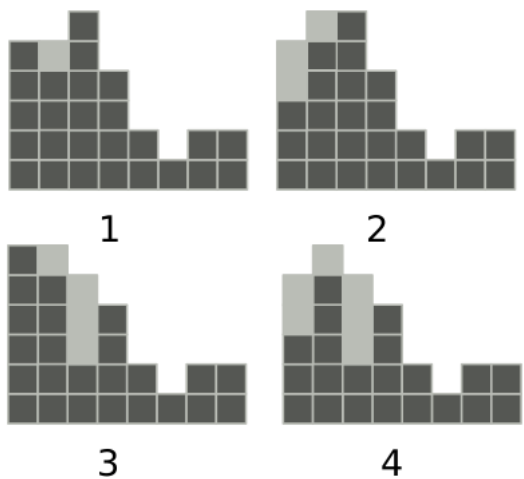}
 \caption{Effects of deposition of a cell in site 2, classified in four cases.}
 \label{neighbors}
\end{figure}

Before applying eq. (\ref{Dlt wq1a}) to the etching model, it is necessary to
know the values of  $\Delta h_{i}(t)$ and $\Delta\overline h(t)$ for one
iteration. While the first may be directly obtained from the model rules, the
second must be separated in four possibilities, depending on the neighbors of
the deposition site $i$. These four situations are shown in Fig.
(\ref{neighbors}) together with the added cells when site $i$ is selected. The
effect of them on $\Delta\bar{h}(t)$ are
\begin{equation}
\Delta\bar{h}(t)=\begin{cases}
\text{Case 1:}& \frac{1}{L} \Delta y.\\
\text{Case 2:}& \frac{1}{L} \left[\Delta y + \left(h_i-h_{i-1}\right)\right].\\
\text{Case 3:}& \frac{1}{L} \left[\Delta y + \left(h_i-h_{i+1}\right)\right].\\
\text{Case 4:}& \frac{1}{L} \left[\Delta y + \left(h_i-h_{i-1}\right) + \left(h_i-h_{i+1}\right)\right].
\end{cases}
\label{deltahm}
\end{equation}

The rate of change of $w_{q}$, eq. (\ref{Dlt wq1a}), can be separated in four
parts corresponding to the cases of eq. (\ref{deltahm}). We represent each of
them as $\left(\frac{\Delta w_{q}}{\Delta t}\right)_{j}$, with $j=[1..4]$. From
eq. (\ref{36c-1}) we conclude that $h_i\sim w$, so, eq. (\ref{Dlt wq1a})
implies that  $\left(\frac{\Delta w_{q}}{\Delta t}\right)_{j}$ is a quadratic
function of $w$. Since the integral (\ref{wqfinal}) is a function of $w$ only
through $\Delta w_q/\Delta t$,  all dependence on $w$ is within that term,
resulting in a quadratic equation for the roughness dynamics,
\begin{equation}
\left< \frac{\Delta w_q}{\Delta t} \right>=-c_a w^2-c_bw-c_c,\label{quadratic}
\end{equation}
where the minus signals were include to simplify forthcoming calculations. In
the limit $\Delta t = 1/L\rightarrow 0$ we can do
\begin{equation}
\left<\frac{\Delta w_q}{\Delta t}\right> \rightarrow \frac{dw_q}{dt} = 2w\frac{dw}{dt}.\label{54c}
\end{equation}
By replacing the Eq. (\ref{quadratic}) in this expression, we obtain
\begin{equation}
-c_a\,dt = \frac{2w\,dw}{w^2+\frac{c_b}{c_a}w+\frac{c_c}{c_a}}
 \equiv \frac{2w\,dw}{(w-w_1)(w-w_2)}
\equiv \frac{A\,dw}{w-w_1}+\frac{B\,dw}{w-w_2},  \label{edo}
\end{equation}
where $w_1$ and $w_2$ are the roots of
$w^2+\frac{c_b}{c_a}w+\frac{c_c}{c_a}=0$, and $A=\frac{2w_1}{w_1-w_2}$, $B=\frac{-2w_2}{w_1-w_2}$.
The solution of the above diferencial equation is
\begin{equation}
\left(\frac{w-w_1}{w_0-w_1}\right)^{A}
\left(\frac{w-w_2}{w_0-w_2}\right)^{B}
= e^{-c_a t}, \label{difsol}
\end{equation}
being $w_{0}$ the initial roughness.

We have found an implicit equation for the roughness as a function of time.
Knowing that that equation and the Family-Vicsek scaling have both the
characteristic time $t_\times$, we conclude that $t_\times = 1/c_a$. We will
show briefly that $w_1 w_2<0$, and, if we choose $w_1>0$, then $w_2<0$, $A>0$
and $B>0$. The limit $w\rightarrow w_s$ when $t\rightarrow \infty$ establishes
$w_1=w_s$. Finally, roughness $w$ depends on $L$ through the constants
$t_\times$ and $w_s$, therefore, if we want the remaining of the equation to be
independent of $L$, we must have $w_2 = -\lambda w_s$ being $\lambda$ a
positive proportionality factor independent of $L$. Incorporating that
reasoning in eq. (\ref{difsol}), and considering a flat initial substrate, we
can write
\begin{equation}
 \left( 1-\frac{w}{w_s}\right)^\frac{2}{1+\lambda}
 \left( 1+\frac{1}{\lambda}\frac{w}{w_s}\right)^\frac{2\lambda}{1+\lambda}
 = e^{-t/t_\times} .
\label{implicit}
\end{equation}

The initial growth, i.e., when $w\ll w_s$, is controlled by the exponent
$\beta$. In order to make explicit that dependence, we expand the left hand
side of Eq. (\ref{implicit}) up to the second order, which result in
\begin{equation}
1 + \frac{1}{\lambda} \frac{w^ 2}{{w_s}^2} + O(3) = e^{-t/t_\times} .
\end{equation}
The inversion of that equation is
\begin{equation}
w(t) \approx w_s \sqrt{\lambda}\left(1-e^{-t/t_\times}\right)^{1/2}.
\label{w12}
\end{equation}
This expression for $w$ is real only if $\lambda>0$, implying $w_1 w_2<0$, as
we said. From the limit of the above expression,
\begin{equation}
\lim_{t\rightarrow 0} w(t) = w_s\sqrt{\lambda} \left(\frac{t}{t_\times}\right)^{1/2},
\label{beta12}
\end{equation}
we find $\beta=1/2$.

It is interesting to make it clear that if $\lambda=1$, Eq. (\ref{w12}) is
identical to Eq. (\ref{implicit}), not an approximation of it.

\subsection{The aftermath of equiprobability assumption}

The exponent $\beta=1/2$ found in the previous section is not the one expected
for the KPZ university class, which we know the etching model belongs to. The
main approximation to be blamed for that disagreement is the equiprobability
assumption, explicitly, that all configurations not forbidden by Eqs.
(\ref{conf}) are allowed, and that they have all the same probability.

\begin{figure}
 \centering
\includegraphics[scale=0.5]{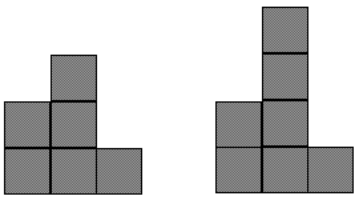}
 \caption{Two possible configurations that can not be generated by the etching model.}
 \label{conf_proib}
\end{figure}

In growth models based in cellular automata, the internal rules of the model
forbid certain configurations. For instance, the algorithm that governs the
etching model can never lead to the configurations shown in the Fig.
\ref{conf_proib}. According to the the model, when one particle is deposited on
the top of the second site, the first and third sites grow up to the earlier
height of the second site, which is not the case of  Fig. \ref{conf_proib}. It
results that these are prohibited configurations.

Beside that, there is no reason to suppose that the dynamics results in equal
probability for the allowed configurations. Indeed, numerical experiments with
the etching model demonstrated that $p(w, h_{i-1}, h_i, h_{i+1})$ doesn't agree
with eq. (\ref{p1}).

The equiprobability assumption disregard the ban of some configurations and the
non-uniformity of the probability distribution, allowing us to express the
probability $p(w, h_{i-1}, h_i, h_{i+1})$ as the ratio of the area of the partial
hypersphere (defined by $w$, $h_{i-1}$, $h_i$, $h_{i+1}$) and the area of the total
hypersphere (defined by $w$). However, the resulting expression for $w$, Eq.
\ref{w12}, has not the expected exponent $\beta$.

Aiming to circumvent the effect of the equiprobability approximations, we write
the \emph{real roughness of the etching model, $w^e$,} as a function of the the
roughness $w$, which does not include those features,
\begin{equation}
w = w_s \left( \frac{w^e}{w^e_s} \right)^\nu,
\label{wew}
\end{equation}
where the superscript refers to saturation. There are good reasons to reduce
all the possibilities to this form. First, $ w$ must increase monotonously with
$w^e$,  they must agree for null roughness, and they have their maximum at
saturation. This is the simplest form that that fulfil those requirements.
Second, for a hypersphere of Euclidean dimension $d$, and radius $\sim R$, the surface
area grows as $R^{d-1}$, while
surfaces such as those originated by etching, or domains as  in phase transition  are 
expected  to grow as $S_f \propto R^{d_f-1}$ , where $d_f$ is the fractal
dimension. I.e., they must have a fractal character and
dimension $d_f$. In this way, Eq. (\ref{wew}) is not only the simplest form, it
is the only possibility.

If we substitute eq. (\ref{wew}) in eq. (\ref{w12}) we can write
\begin{equation}
w^e(t)=w_s^e\left[1-\exp\left(-\frac{t}{t_\times}\right)\right]^{\beta}.
\label{wet}
\end{equation}
where $\beta=\frac{1}{2\nu}$. If we explicitly make $w_s^e \propto L^\alpha$
and $t_\times \propto L^z$, that expression becomes one possible form of
Family-Vicseck scaling relation, Eq. (\ref{FamilyVicsek}).

\begin{figure}
 \centering
\includegraphics[scale=1.0]{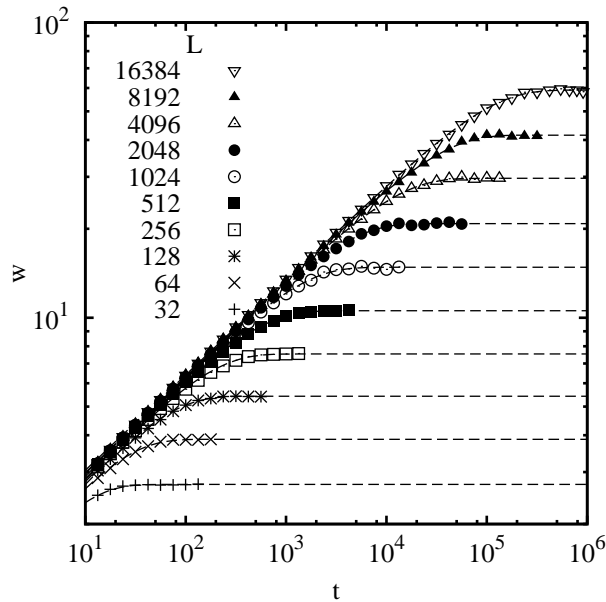}
 \caption{Roughness as a function of the time.
 The points are the results of numerical simulation of etching model for several substrat lenghts.
 The curves are fittings of Eq. (\ref{wet}) to the data points.}
 \label{fig_1d}
\end{figure}

Although that equation is an expansion strictly valid only for $t\ll t_\times$
it nicely fit to the results of the numerical simulations in the whole range of
$t$, as can be seen in Fig. (\ref{fig_1d}). Each curves of that figure was
obtained by fitting its parameters $\beta$, $t_\times$, and $w_s$ to the
corresponding data points. Data from simulation of other surface evolution
models (RSOS, Edwards-Wilkinson) and higher dimensions have shown the same
exceptional agreement with Eq. (\ref{wet}).

\section{Conclusions}

In this work, we presented a method to obtain the roughness equation of the
models. The method is based on the ratio of the hypersphere areas, interpreted
as the probability of occurrence of determined configurations of the interface.
The hyperspheres method has potential of application in automata cellular
models which depend only on the nearest neighbors  but it needs to be built
differently for each type of algorithm. The algorithms need to act only in the
nearest neighbors and the systems need to be one-dimensional. If approximations
similar to equiprobability assumption are necessary to solve these models,
transformation similar to Eq. (\ref{wew}) may be necessary.

The equation possesses three parameters, each of them is connected with one of
the growth exponents. The modified equation is relevant not only because it
fits well to the data, but also because we can also be used as fitting function
for obtaining the value of the main parameters of the model, $w_s^e$, $\beta$,
and $t_\times$.  In correlated stochastic phenomena , analytical results are rather difficult to obtain. In this way we hope that  this work may inspire research into those systems where even not exact solutions can be considered major results \cite{Ferreira12}.

\textbf{Acknowledgements:} This work was supported by CAPES, CNPq, and FAPDF.

\end{document}